\documentclass[smallcondensed]{svjour2}

\smartqed  
\usepackage{color}
\usepackage{graphicx}
\usepackage{mathptmx}      
\usepackage{amsmath}
\usepackage{amssymb}
\usepackage{times}
\usepackage{epsfig}
\usepackage{psfrag}

\journalname{J Stat Phys}
\begin{document}
\title{ 
Finite size corrections to  the large deviation function of the density in the
one dimensional symmetric simple exclusion process
}

\author{Bernard Derrida \and Martin Retaux}

\institute{   B. Derrida \and M. Retaux \\
              Laboratoire de Physique Statistique,\\
              \'Ecole Normale Sup\'erieure, 
 Universit\'e Pierre et Marie Curie, Universit\'e Denis Diderot, CNRS \\
              24, rue Lhomond,
              75231 Paris Cedex 05 - France \\
              \email{derrida@lps.ens.fr}\and\email{martin.retaux@ens.fr}
}

\date{Received: date / Accepted: date}

\maketitle

\begin{abstract}
{The symmetric simple exclusion process is one of the simplest out-of-equilibrium systems for which the steady state is known. Its large deviation functional 
 of the density has been computed in the past both by microscopic and  macroscopic approaches.
Here we obtain the leading finite size correction to this large deviation functional. The result is compared to the similar corrections for equilibrium systems.
}
\end{abstract}
\ \\ \ \\
\today \\ \ \\
\section{Introduction}

Over recent years there has been a growing interest in understanding the fluctuations
 and the large deviations of the density of  systems in  a non equilibrium  steady state \cite{KOV,BDGJL1,BDGJL2,BDGJL3,BDGJL4,BDGJL5,BDGJL6,DLS2001,DLS2002a,DLS2002b,DLS2003,enaud,derrida-2007,RV,BGL,WZ,GKP,LT,GKLT,Touchette,haifa1}.
In such steady states, the generic situation is that  the   correlation range of  density fluctuations 
 extends through the whole system  \cite{KCD,spohn-cor,SC,DKS,OS,DLS2007,BDLW,bertini-lr} and the large deviation functional  of the density  is non local \cite{BDGJL1,BDGJL2,BDGJL3,BDGJL4,BDGJL5,DLS2001,DLS2002a,DLS2002b,DLS2003,enaud,derrida-2007}.  
This  contrasts with     systems (with short range interactions and far from a critical point) at equilibrium,  where  the range of correlations 
is microscopic  and the large deviation functional   is  local. 

Two main approaches have been followed recently to study these large deviations: for some microscopic models such as exclusion processes the steady state measure is known exactly  \cite{DLS2001,DLS2002a,DLS2002b,DLS2003,enaud,derrida-2007} and finding the large deviation functional is a matter of computing large scale properties (very much like when one tries to calculate the free energy in equilibrium systems starting from the Gibbs measure).
Obviously this microscopic approach is limited to cases where the steady state is exactly known.
The other approach is the macroscopic fluctuation theory \cite{BDGJL1,BDGJL2,BDGJL3,BDGJL4,BDGJL5,BDGJL6}) for diffusive systems which calculates the large deviation functional by identifying the optimal path followed by the system to generate a given deviation. In  systems at equilibrium, time reversal symmetry gives a simple relation between this path and the relaxation path starting from the same deviation,   and so identifying this optimal path is easy. On the contrary in non equilibrium steady states this time reversal symmetry does not hold and  the  { approach is limited to cases where  the equations giving this optimal path
can be solved}.

One motivation to study  the large deviation functional of the density is that it generalizes the notion of free energy  to  non equilibrium states \cite{BDGJL6,derrida-2007}.
 As analytic expressions of these large deviation functionals are usually  hard to obtain,   they   are known so far 
 for a rather limited number of models.
The one dimensional symmetric simple exclusion process (SSEP)  was one of   the   first   models  \cite{BDGJL2,BDGJL5,DLS2001,DLS2002a,derrida-2007} for which an explicit expression
 could be derived which showed the non local character 
 of this  large deviation functional.
The goal of the present work is to obtain the leading  finite size corrections to this large deviation functional and to compare it with the corrections one typically finds in equilibrium systems.

 The SSEP   describes  a lattice of $L$ sites in which each site   $i$ is either occupied  by a single particle or empty \cite{spitzer,E1,E2,spohn-book,liggett,kipnis-landim}.
Each particle independently    attempts  to jump to its right neighboring site, and to its left
neighboring site with rate one. It succeeds if the target site is empty; otherwise nothing happens. 
At
the boundary sites, 1 and $L$, particles are added or removed: a particle is
added to site 1, when the site is empty, at rate $\alpha$, and removed, when the
site is occupied, at rate $\gamma$; similarly particles are added to site $L$ at rate $\delta$  
and removed at rate $\beta$.
These injection and removal rates at the boundaries correspond to the left and right boundaries being in contact with reservoirs at densities
\begin{equation} 
\rho_a= {\alpha \over \alpha + \gamma} \ \ \ ; \ \ \ 
\rho_b= {\delta \over \beta + \delta}
\label{ra-rb}
\end{equation} 
(to justify (\ref{ra-rb}) it is easy to check using detailed balance   that,  when one forbids the exchanges of particles at site $L$ by setting $\beta=\delta=0$,  the steady state measure is  a Bernoulli measure where all
the sites are occupied with probability $\rho_a$. Similarly one can check that  when the contacts between site  $1$  and the left reservoir are broken, the system equilibrates at density $\rho_b$.)
The main advantage of the SSEP is that its steady state measure is known  \cite{spohn-cor,DEHP,DLS2001,DLS2002a} for arbitrary $\alpha,\beta,\gamma,\delta$ and $L$.

Here we  try to determine the generating function of the density, which is simply the Legendre transform of the large deviation  functional.
Let $P(n_1, \cdots, n_L)$ be the steady state measure  of a
  one dimensional
lattice gas   on a lattice of $L$ sites,  where  $n_i \ge 0$ is the   number of particles on site $i$
(for the SSEP the only possible values are $n_i=0$ or $1$ but in the   more general case  discussed in  section 4 the occupation numbers 
$n_i$ will be arbitrary).
We want  to calculate the  following generating function  $Z_L(h_1, \cdots, h_L) $ (which, in the equilibrium case,  is  nothing but a partition function in a varying field)
\begin{equation}
Z_L(h_1, \cdots, h_L) =  
\sum_{\{ n_1, \cdots n_i \} } \exp\left( \sum_{i=1}^L h_i  \,n_i \right) \ P(n_1, \cdots n_L) 
\label{Z-def}
\end{equation}
 where $h_i$ depends on the site $i$.
Let  us define $G_L(h_1, \cdots h_L)$ as
\begin{equation}
G_L(h_1, \cdots h_L)  =  \log Z_L(h_1, \cdots, h_L)  \ .
\label{G-def}
\end{equation}
We would like to obtain an expression of $G_L(h_1, \cdots, h_L)$ for a slowly varying field, that is when $h_i$ is of the form
\begin{equation}
h_i = H\left({i \over \lambda }  - {1 \over 2 \lambda} \right)
\label{slow}
\end{equation}
where $\lambda$ is large 
(the reason for the shift of $-1/2 \lambda$ in   (\ref{slow}) is simply to make sites $1$ and $L$ play symmetric roles).
This choice for the $i$  dependence of $h_i$ allowes one to test density deviations which vary on  a  large  length scale $\lambda$, which
might be different  from the system size $L$

In the following  we will consider the case where the two lengths  $\lambda$ and $L$ are large  (compared with the lattice spacing) but  comparable 
\begin{equation}
 L =  \lambda   y   
\label{y-def}
\end{equation}
with $y$ of order 1.
 For the symmetric exclusion  process
in contact at site $i=1$ and at site $i=L$ with two reservoirs at densities
$\rho_a$ and $\rho_b$ it is known \cite{DLS2001,DLS2002a,BDGJL2,BDGJL3}  (see for example  eq. (80,81)  of \cite{derrida-2007})  that, in  the steady state, 
\begin{equation}
G_L(h_1,\cdots h_L)\sim \lambda \min_{\{F(x) \}} \int_0^y dx \left( \log(1  + F(x) (e^{H(x)}-1) ) - \log \left( y \, {F'(x) \over \rho_b- \rho_a} \right)  \right)
\label{G-ext}
\end{equation}
where  the minimum is over all the monotone functions   $F(x)$
  which satisfy $F(0)=\rho_a$ and $F(1)= \rho_b$. 
From (\ref{G-ext}) it is easy to see that the equation satisfied by the optimal $F$ is  
\begin{equation}
F''(x) = {F'(x)^2  \, (e^{H(x)} -1) \over 1 - F(x) + F(x)  \, e^{H(x)}} 
\label{Feq}
\end{equation}
Note that the non local character of the functional (\ref{G-ext}) comes from the fact that  the optimal  $F(x)$ depends on all values of $H(z) $ for  the whole range $0 < z <y= {L \over \lambda} $.

On the other hand for a system at equilibrium (with short range interactions)  one expects \cite{BDGJL6,derrida-2007} that
\begin{equation}
G_L(h_1,\cdots h_L)\sim \lambda \int_0^y dx \ g(H(x))
\label{Geq}
\end{equation}
 where $g(h) = \lim_{L \to \infty} {G_L(h,\cdots,h)  / L}$ is the extensive part of the free energy in a constant $h$.
This is obviously a local functional of $H(x)$.
\ \\ \ \\
The main result presented in the present  work is that the leading correction to
(\ref{G-ext}) is 
\begin{eqnarray}
G_L(h_1,\cdots h_L)\simeq &    \lambda \int_0^y dx \left( \log(1  + F(x) (e^{H(x)}-1) ) - \log \left({ y \, F'(x) \over \rho_b- \rho_a} \right)  \right)&
\nonumber \\
 & - a \log\left( {y \, F'(0) \over   \rho_b-\rho_a} \right)
- b \log\left( {y   \, F'(y   ) \over  \rho_b-\rho_a} \right)
-{1 \over 2} \log( \varphi(0))&
\label{res-neq}
\end{eqnarray}
where $F$ is the solution of (\ref{Feq}), the parameters $a$ and $b$ are defined as in \cite{DLS2002a,DLS2007}
\begin{equation}
a={1 \over \alpha + \gamma}  \ \ \ ; \ \ \  b={1 \over \beta+ \delta}
\label{a-b-def}
\end{equation}
and $\varphi(x)$ is the solution of the linear differential equation 
\begin{equation}
\varphi''(x) =  -\left({F''(x) \over F'(x) }  \right)' \varphi(x) 
\label{phieq}
\end{equation}
which satisfies the boundary conditions $\varphi(y)=0 $ and  $ \varphi'(y)=-1/y$.

This can be compared to the case of a systems at equilibrium where  the form of the leading orders of $G_L$ is
\begin{eqnarray}
 G_L(h_1, \cdots h_L) \simeq  \lambda  \int_0^{y}  g( H(x))  \ dx  \ +  A \ ^{\rm left} (H(0))  \ +  \ A^{\rm right} \left(H\left( y\right)\right)  &&
\nonumber \\ 
  +  \nonumber 
{1 \over \lambda}  \left[  H'(0)  \ B^{\rm left}   (H(0))
+ H'\left(  y \right) \    B^{\rm right} \left(H\left(y \right)\right)   \right. &&
 \\   \left. + \int_0^{y}  C( H(x))\  H'(x)^2 \  dx  \right]    + 0 \left( {1 \over \lambda^2 } \right) &&
 \label{res-eq}
\end{eqnarray}
where $A^{\rm left} (h), A^{\rm right} (h), B^{\rm left} (h), B^{\rm right} (h) $ and $C(h)$ are defined in (\ref{G-asymp},\ref{B-def},\ref{C-def}).
We see that the leading correction  (i.e. the term of order $0$ in $\lambda$)  is also non local (\ref{res-neq})  in the out of equilibrium SSEP 
whereas it corresponds to boundary contributions  $ A^{\rm left} $ and $ A^{\rm right}$ in the equilibrium case (\ref{res-eq}). At the next order  (the order $1/\lambda$),  which we did not study in the non-equilibrium case, 
 one can notice  in (\ref{res-eq})
an integral containing  the    gradient term   $H'(x)^2$ characteristic of the Ginzburg-Landau theory.
 \ \\ \ \\ 
The paper is organized as follows. In section 2, we do a direct perturbative calculation  when the $h_i$'s are small and we check that the expansion agrees with
the  prediction (\ref{res-neq}) for the SSEP. In section 3, we present the derivation of (\ref{res-neq})  for arbitrary $h_i$'s. In section 4, we 
discuss  how (\ref{res-eq})  can be derived in the equilibrium case.
\section{Perturbations for  small $h_i$}
In this section we present the straighforward calculation of $G_L$ from the knowledge of the correlation functions of the density in the steady state.
From the definition (\ref{G-def}), one can relate the expansion of $G_L$ in powers, of the $h_i$'s to the steady state correlations. For example to second order in the $h_i$'s
 one has
\begin{equation}
G_L(h_1, \cdots h_L)    = 
\sum_i h_i \,  \langle n_i\rangle  + 
\sum_i  
{h_i^2  \over 2}  (\langle n_i^2  \rangle - \langle n_i  \rangle^2) + 
 \sum_{i < j}   h_i h_j  \ \langle n_i n_j \rangle_c 
+ O\left(h^3 \right)  
\label{G-exp1}
\end{equation}
For the SSEP, it is known that in the steady state  \cite{spohn-cor,DLS2007} 
\begin{eqnarray}
\label{ni}
  \langle n_i\rangle =   \langle n_i^2\rangle
  &=&{\rho_a(L+b-i)+\rho_b(i+a-1)\over L+a+b-1},\\
  \langle n_i n _j\rangle_c
   &=&-{(\rho_a-\rho_b)^2(i+a-1)(L+b-j)\over(L+a+b-1)^2(L+a+b-2)} \ \ \ \ \ \ \ {\rm for} \ \ \  i < j
\nonumber
\end{eqnarray}

For large $L$ and $\lambda$ (keeping their ratio constant as in (\ref{y-def}), when the $h_i$ have the form (\ref{slow}) the various sums in (\ref{G-exp1}) can be computed by the Euler Mac Laurin formulae:
$$ \epsilon \sum_{i=1}^L f\left(i \epsilon-{\epsilon \over 2} \right) = \int_0^{L \epsilon} f(x) dx -    \epsilon^2 {[f'(L \epsilon) - f'(0)] \over 24} + 
 O\left(\epsilon^4\right) $$
 \begin{small}
 \begin{equation*}
 \epsilon^2 \sum_{1 \leq i < j \leq L} f\left({i \epsilon } - {\epsilon \over 2}  \right) g\left(j   \epsilon - {\epsilon \over 2} \right)
 =   \int_0^{L \epsilon}  f(x) dx  \int_x^{L \epsilon}  g(y) dy  
 \ - \  {\epsilon  \over 2} \int_0^{L \epsilon}   f(x) g(x)  dx  + O(\epsilon^2)  \  
\end{equation*}   
\end{small} 
and one gets 
\begin{small}
\begin{eqnarray}
\label{linear}
 \sum_i  h_i \langle n_i \rangle 
&& = \lambda \int_0^y \overline{\rho}(x) H(x) dx   + (\rho_a- \rho_b) \int_0^y {y  - 2 a y  - 2 x + 2 a x + 2 b x  \over 2 y^2} H(x) dx \\
\nonumber \sum_i  h_i^2 \langle n_i  \rangle 
&&= \lambda \int_0^y \overline{\rho}(x) H^2(x) dx   + (\rho_a- \rho_b) \int_0^y {y  - 2 a y  -2 x + 2 a x + 2 b x  \over 2 y^2} H^2(x) dx \\ 
\nonumber \sum_i  h_i^2 \langle n_i  \rangle ^2
&&= \lambda \int_0^y \overline{\rho}^2(x) H^2(x) dx   + (\rho_a- \rho_b) \int_0^y {y - 2 a y -2 x +2 a  x   + 2 b x \over  y^2}  \overline{\rho}(x) H^2(x) dx
\end{eqnarray}
\end{small} 
 where  $\overline{\rho}(x)$ is the steady state profile
$$\overline{\rho}(x) = {\rho_a (y-x) + \rho_b x \over y} $$
and
\begin{eqnarray*}
 \sum_{i<j} h_i h_j  \langle n_i   n_j\rangle_c && =  -\lambda  (\rho_a-\rho_b)^2 \int_0^y dx \int_x^y dz {x (y-z) \over y^3}  H(x) H(z)  \\
&& + {(\rho_a- \rho_b)^2 \over 2 y^4 } \left[ y \int_0^y x (y-x) H(x)^2 dx + \int_0^y dx \int_x^y dz \  H(x) H(z) \times  \right. \\
&& \left. \Big[(1-2 a ) (y-x) (y-z)+ (4 a  + 4 b -6)  x (y-z)   + (1- 2 b)  x z \Big]  \right]
\end{eqnarray*}
We have checked that these expressions coincide with (\ref{res-neq})  at  second order in $H(x)$.
 For example at first order in $H(x)$ the solutions of  (\ref{Feq}) and (\ref{phieq}) are
$$F(x)= {\rho_a (y-x) + \rho_b x \over y} + {(\rho_a-\rho_b)^2  } \left[  \int_x^y {x (z-y) \over y^3} H(z) dz  -   \int_0^x {  z (y-x) \over y^3}H(z) dz \right]  $$
$$\varphi(x)= (y-x)/y  + {(\rho_a-\rho_b) \over y^2} \int_x^y (2 z - y - x) H(z) dz 
$$
and inserting these expressions into (\ref{res-neq}) one gets (\ref{linear}).

\section{Derivation of  the main result (\ref{res-neq})} 
Our approach  to obtain (\ref{res-neq}) consists in choosing  $h_i$  piecewise constant: $h_i$ takes $n$ possible values $H_1, \cdots H_n$ in $n$ consecutive boxes. As
in each of these  boxes, $h_i$ is constant we will use the expression (\ref{Z1box},\ref{mu0})  for a single box with a constant $h$ which is much easier to obtain.
Then we will use an additivity formula  (\ref{additivity5})  to go from  the expression for  one box to the expression for $n$ boxes. Finally we will take the limit $n \to \infty$
to establish (\ref{res-neq}).
\subsection{\bf  A single box}
Using the matrix ansatz (see the Appendix \ref{annexe1}), one  gets, for large $L$,  the following expression for $Z_L(h,h, \cdots h)$  by dividing 
(\ref{1box-large-L}) by (\ref{normalization})
\begin{eqnarray}
\label{Z1box}
 Z_{L}(h,\cdots,h)  \simeq    {  (\rho_a-\rho_b)^{L+a+b}
 \ \mu_0^{-L-a-b}  
 \over  (1+ \rho_a(e^h-1))^{a}\, (1 +  \rho_b(e^h-1))^{b} }\, 
\end{eqnarray}
where
\begin{equation}
\mu_0= {1 \over  e^h -1   }\log { 1 + \rho_a (e^h    -1) \over 1 + \rho_b (e^h-1)} 
\label{mu0}
\end{equation}
{\bf Remark:}
Let us check that these expressions  agree with the claim (\ref{res-neq}) in the introduction:
 one has by solving (\ref{Feq}) and (\ref{phieq})  for a constant $h$ 
$$F(x)= {1 \over e^h-1}\left[( 1 + \rho_a(e^h-1))^{{1-x  / y}} \, (1+\rho_b(e^h-1))^{x/y} \,  -  \, 1 \right] $$
and $$\varphi(x) = {y-x \over y}$$
from which it follows that
$$\log [1 + (e^h-1) F(x) ] - \log\left({y \,  F'(x) \over \rho_b- \rho_a}\right) = \log \left({\rho_a-\rho_b \over \mu_0}\right)$$
$$F'(0)= {1 + \rho_a(e^h-1) \over y (e^h-1)} \log\left({1+\rho_b(e^h-1) \over 1 + \rho_a(e^h-1)} \right)$$
$$F'(y)= {1 + \rho_b(e^h-1) \over y (e^h-1)} \log\left({1+\rho_b(e^h-1) \over 1 + \rho_a(e^h-1)} \right)$$
and by replacing into (\ref{res-neq})  one  finds 
an  expression  equivalent to  (\ref{Z1box}) obtained by the direct calculation.  This shows that (\ref{res-neq}) is  valid in the case of a constant $h_i$.
\subsection{\bf Several  boxes: the prediction (\ref{res-neq})}
Let us now come to the case of several large boxes
with a constant $h_i$ in each box. We will first   write down the expressions predicted by the claim (\ref{res-neq}). Then we will see in the next subsection that these expressions coincide with those obtained by a direct microscopic calculation.

For  piecewise constant $H(x)$, with 
\begin{equation}
\label{Hxm}
  H(x) = H_m  \ \ \  {\rm for  } \ \ \  x_{m-1}  <x < x_m
\end{equation}
with
\begin{equation}
x_0=0 \ \ \ \ ; \ \ \ \ x_m=x_{m-1}+ y_m \ \ \ ; \ \ \  x_n=y
\label{xm}
\end{equation}
 the solution of (\ref{Feq}) 
in the interval  $x_{m-1}<x<x_m$ is 
\begin{equation}
\label{Fx-sev}
  F(x) = { 1 \over e^{H_m}-1}
\left[(1+ (e^{H_m}-1)F_{m-1})^{x_{m}-x \over x_{m}-x_{m-1}}  (1+ (e^{H_m}-1)F_{m})^{x-x_{m-1} \over x_{m}-x_{m-1}} -1  \right]
\end{equation}
where $F_m= F(x_m)$.
Writing  that $F'(x)$ is continuous  (i.e. $F'(x_m)^-=F'(x_m)^+$) at all the $x_m$'s  leads to the $n-1$ equations  that these $F_m$'s should satisfy
\begin{small}
\begin{equation}
{ 1 + (e^{H_m} -1)F_m \over  y_m(e^{H_m} -1)} \log\left( {1 + (e^{H_m}-1) F_{m} \over 1 + (e^{H_m}-1) F_{m-1}} \right)
=
{ 1 + (e^{H_{m+1}} -1)F_m \over  y_{m+1}(e^{H_{m+1}} -1)} \log \left({1 + (e^{H_{m+1}}-1) F_{m+1} \over 1 + (e^{H_{m+1}}-1) F_{m}} \right)
\label{Fm-eq}
\end{equation}
\end{small}
Equations (\ref{Fx-sev},\ref{Fm-eq}) fully determine the solution of (\ref{Feq}) for a piecewise constant $H(x)$.

To solve the equation (\ref{phieq}) for $\varphi(x)$, one can first notice that     the discontinuity of the $\varphi'(x)$ at $x=x_m$ is 
$$\varphi'(x_m)^+ - \varphi'(x_m)^- =  \, { F''(x_m)^- - F''(x_m)^+  \over F'(x_m)} \, \varphi(x_m) \ . $$
Everywhere else   the function $\varphi(x)$ is piecewise linear.  These jumps of $\varphi'(x)$ and the fact that $\varphi(y)=0$ and $\varphi'(y)=-1/y$
determine the function $\varphi(x)$  everywhere: 
in the interval $x_m < x < x_{m+1}$ one gets
\begin{small}
\begin{eqnarray}
\label{phisol}
 \varphi(x) = {1 \over y} \Big[  y-x   && +  \sum_{m_1 > m}  (y-x_{m_1})( x_{m_1}-x) \, W_{m_1}  
\\
  && +  \sum_{m_1 > m_2 > m} (y-x_{m_1})( x_{m_1}-x_{m_2})(x_{m_2}-x ) W_{m_1} \, W_{m_2}  
\nonumber
\\
   && +   \sum_{m_1 > m_2 > m_3> m}  {(y-x_{m_1})( x_{m_1}-x_{m_2})(x_{m_2}-x_{m_3} )(x_{m_3}-x ) }\, W_{m_1} \, W_{m_2}  \, W_{m_3}  + ... \Big]
\nonumber
\end{eqnarray} \end{small} 
with  $W_m$ defined by 
\begin{equation}
W_m={ F''(x_m)^- - F''(x_m)^+  \over F'(x_m)}  \ . 
\label{Wm-def}
 \end{equation}
Using (\ref{Fx-sev}) and (\ref{Fm-eq}), one can show that
\begin{equation}
W_m= {1 \over y_m} \log\left[{1+ (e^{H_{m}} -1) F_m  \over 1+ (e^{H_m} -1) F_{m-1} } \right] -  
{1 \over y_{m+1}} \log\left[{1+ (e^{H_{m+1}} -1) F_{m+1}  \over 1+ (e^{H_{m+1}} -1) F_{m} } \right]
\label{Wm-exp}
 \end{equation}

In summary in the case of several large boxes the claim (\ref{res-neq}) leads to
\begin{equation}
Z_L(h_1, \cdots, h_L) =
e^{G_L(h_1, \cdots h_L) }  \simeq {\cal B} \, \exp[\lambda {\cal C}(\rho_a,\rho_b)] 
\label{claim1}
\end{equation}
with
\begin{equation}
{\cal C} = -\sum_m y_m  \, \log \left[{ y \over y_m (\rho_b-\rho_a) (e^{H_m}-1) } \log \left( {1 + ( e^{H_m}-1) F_m \over 1 + ( e^{H_m}-1) F_{m-1}} \right) \right]
\label{claim2}
\end{equation}
and
\begin{equation}
{\cal B}= 
\left({ \rho_b- \rho_a \over y \, F'(0)}\right)^a 
 \ \left({ \rho_b- \rho_a \over y \, F'(y)}\right)^b \, {1 \over \varphi(0)^{1/2}} 
\label{claim3}
\end{equation}
with $F(x)$ and $\varphi(x)$  given by (\ref{Fx-sev},\ref{phisol}) and the $F_m$'s  solutions of (\ref{Fm-eq}).

\subsection{\bf Several  boxes: the microscopic approach}
Let us now see how the microscopic calculation for the single box can be generalized to the case of several boxes and leads to expressions equivalent to (\ref{claim2},\ref{claim3}).
We consider   the case of several large boxes
with a constant $h_i$ in each box 
\begin{equation}
h_i= H_m \ \ \ \ \ \ \ \  \ {\rm  for} \ \ \   \ \ L_1+ \cdots + L_{m-1}  < i \leq L_1+ \cdots + L_m
\label{Hn}
\end{equation}
  uses the additivity formula (\ref{additivity5}) and the saddle point method.

Let us define 
$$z_1(\rho_a,\rho) = 
{ \langle \rho_a,a |    \left(e^{H_1} D + E\right)^{L_1}     | \rho,b \rangle  \over \langle\rho_a,a  |    (D + E) ^{L_1}    |\rho,b \rangle }
$$
and for $i  \ge 2$
$$z_i(\rho,\rho') =
{ \langle \rho,1-b |    \left(e^{H_i} D + E\right)^{L_i}     | \rho',b \rangle  \over \langle\rho,1-b  |    (D + E) ^{L_i}    |\rho',b \rangle }
$$ which are the generating functions for each box (see Appendix \ref{annexe1}).
($z_1$ is special  simply because in $z_2, \cdots z_n$ the parameter $a$ has been replaced by $1-b$).
Then using the additivity formula (\ref{additivity5}) derived in  Appendix \ref{annexe1}  one gets for
 $Z_L(h_1, \cdots,h_L)$
when
$$L_1+L_2 + \cdots L_n=L$$
and the $h_i$ are of the form  (\ref{Hn}) 
\begin{eqnarray}
\label{int-mult}
 Z_L(h_1, \cdots,h_L) =  {\Gamma(L_1+a+b) \, \Gamma(L_2+1) \cdots \Gamma(L_n+1) \over \Gamma(L+a+b)} 
\times 
\ \ \ \ \
\\
\oint   \frac{d\rho_1}{2i\pi}
\cdots \oint   \frac{d\rho_{n-1}}{2i\pi} 
{(\rho_a- \rho_b)^{L+a+b} 
\ z_1(\rho_a,\rho_1) \  z_2(\rho_1,\rho_2) \cdots  \, z_n(\rho_{n-1},\rho_b) 
\over 
\ \ \ \ \ \ (\rho_a- \rho_1)^{L_1+a+b} 
(\rho_1- \rho_2)^{L_2+1}  \cdots (\rho_{n-1} - \rho_b)^{L_n+1} }
\nonumber
\ \ \ \ \
\end{eqnarray}
where the integral contours verify $\rho_b < |\rho_{n-1}| < \cdots <  |\rho_1|  < \rho_a$. 
So far (\ref{int-mult}) is exact for arbitrary $H_m$'s
and $L_m$'s. The virtue of (\ref{additivity5}) is that it relates the properties of the whole system of those of the $n$ subsystems.
\ \\ \ \\
When the lengths $L_m$ of the boxes become large, if we define the $y_m$'s by
$$L_m= \lambda  \, y_m$$
 one knows (\ref{Z1box}) from the single box calculation  that
$$z_m(\rho,\rho') \sim B_m(\rho, \rho')\, e^{\lambda \, y_m \, [A_m(\rho,\rho') + \log(\rho-\rho')]}  \ , $$
with 
\begin{equation}
A_m(\rho,\rho') = - \log\left[
 {1 \over  e^{H_m} -1   }\log { 1 + \rho (e^{H_m}    -1) \over 1 + \rho' (e^{H_m}-1)}\right] 
\label{Am-def}
\end{equation}

\begin{equation}
B_1(\rho,\rho') =
{(\rho-\rho')^{a+b}
\, e^{(a+b) A_1(\rho,\rho') }
\over  (1+ \rho(e^{H_1}-1))^{a}\, (1 +  \rho'(e^{H_1}-1))^{b} 
}
\label{B1-def}
\end{equation}
and for $i \ge 2$
\begin{equation}
B_i(\rho,\rho')=
{(\rho-\rho')\, e^{ A_i(\rho,\rho')} \over (1+ \rho(e^{H_i}-1))^{1-b}\, (1 +  \rho'(e^{H_i}-1))^{b} 
 }
\label{Bi-def}
\end{equation}
Then   using the saddle point method in (\ref{int-mult}) one finds that

\begin{equation}
\label{ZL-star}
Z_L(h_1, \cdots,h_L) 
 \simeq  {\cal B }^*  \  e^{\lambda  [{\cal A}^*+  y \, \log(\rho_a-\rho_b) ]}
\end{equation}
where 
\begin{equation}
\label{calA}
{\cal A}^* = \min_{\{r_m \}}\left[ y_1 A_1(\rho_a,r_1) +  y_2 A_2 (r_1,r_2) + ... + y_n A_n(r_{n-1},\rho_b) \right] -y \log y + \sum_{i=1}^n{y_i \log y_i} \ , 
\end{equation}
\begin{equation}
\label{calB}
{\cal B}^* \simeq   { (\rho_a- \rho_b)^{a+b}   \over y^{a+b-{1 \over 2}} }\  {y_1^{a+b-{1 \over 2}}    B_1 (\rho_a,r_1) \over (\rho_a- r_1)^{a+b}}  \   {y_2^{{1 \over 2}} B_2 (r_1,r_2) \over r_1 - r_2} ...
{y_n^{{1 \over 2}} B_n(r_{n-1},\rho_b)
 \over r_{n-1} -  \rho_b}  \ \left(\det[\Delta]\right)^{-1/2}
\end{equation}
 and  $\Delta$ is a tridiagonal matrix
\begin{equation}
\Delta=\left( \begin{array}{cccccc} U_1   &  V_1  & 0  &0 & 0  &  0\\    V_1 &  U_2   &   V_2 & 0  &0  & 0  \\  0 & . &  . & . &0  & 0   \\ 
 0 & 0 & . & .  & .  & 0 \\
 0 & 0 & 0 & .  & .  & . \\
0&  0 & 0 & 0 &   V_{n-2} &  U_{n-1} \end{array} \right) 
\label{Delta-def}
\end{equation}
with
\begin{equation}
U_m = \left. {\partial^2 
\left[
y_{m} \, A_m(r_{m-1}, \rho) + 
y_{m+1}\,  A_{m+1}(\rho,r_{m+1}) \right]
\over \partial \rho^2 }
 \right|_{\rho=r_m} 
\label{Um-def}
\end{equation}
\begin{equation}
V_m = \left. {\partial^2 
[y_{m+1}\,  A_{m+1}(\rho, \rho')  ]
\over \partial \rho  \, \partial \rho'}
 \right|_{\rho=r_m \, , \,  \rho'=r_{m+1}} 
\label{Vm-def}
\end{equation}
(Note that in (\ref{calA}) one takes the minimum and not the maximum because the integration contours are perpendicular to the real axis, and  a maximum  over $r_m$ along a contour becomes a minimum when $r_m$ varies along the real axis).
\medskip

The saddle point values $r_1,  \cdots r_{n-1}$ (i.e. those which achieve the minimum in (\ref{calA}) )
satisfy the $n-1$ equations
$$
\left. {\partial
\left[
y_{m} \, A_m(r_{m-1}, \rho) +
y_{m+1}\,  A_{m+1}(\rho,r_{m+1}) \right]
\over \partial \rho }
 \right|_{\rho=r_m}  =0
$$
These saddle point equations  turn out to be  the same  equations   as those 
satisfied 
(\ref{Fm-eq}) 
satisfied by the $F_m$'s. Therefore one has
\begin{equation}
\label{rm-Fm}
r_m=F_m
\end{equation}
This already allows one to verify, using (\ref{Am-def},\ref{calA}),  that the term proportional to $\lambda$ in (\ref{claim1}) and (\ref{ZL-star}) is the same.
\\ \ \\
\ \\
One can also show by a direct computation that the $U_m$'s and the $V_m$'s defined 
in   (\ref{Um-def},\ref{Vm-def})   can be expressed in terms of  the function $F(x)$ given in 
 (\ref{Fx-sev}).
\begin{equation}
U_m=\left( {1 \over y_m} + {1 \over y_{m+1}} \right) {1 \over F'(x_m)^2} + {F''(x_m) ^{(-)}  -F''(x_m)^{(+)} \over F'(x_m)^3 }
\label{Umsol}
\end{equation}
\begin{equation}
V_m=-{1 \over y_{m+1}  F'(x_m) F'(x_{m+1})}  \ .
\label{Vmsol}
\end{equation}

It is then easy to see that one can  rewrite $U_m$ as
$$U_m=\left( {1 \over y_m} + {1 \over y_{m+1}} \right) {1 \over F'(x_m)^2} + 
{W_m \over  F'(x_m)^2 } $$
 with $W_m$ given in (\ref{Wm-def}).
Then the determinant of the matrix $\Delta$  defined in (\ref{Delta-def}) can be computed
\begin{eqnarray}
\det[\Delta]= && {1\over y_1  \cdots  y_n [ F'(x_1) \cdots  F'(x_{n-1})]^2}   \   \times    \\
&& \Big[ y  + \sum_m (y-x_m) \,  x_m \, W_m  +   \sum_{m_1 > m_2 } (y-x_{m_1})( x_{m_1}-x_{m_2})x_{m_2} W_{m_1} \, W_{m_2}  \nonumber \\  
 && \ \ \ \ \ + \sum_{m_1 > m_2 > m_3} {(y-x_{m_1})( x_{m_1}-x_{m_2})(x_{m_2}-x_{m_3} )x_{m_3} }\, W_{m_1} \, W_{m_2}  \, W_{m_3}  +  ... \Big]
\nonumber
\end{eqnarray} 

and  using the fact  (see (\ref{Am-def},\ref{Fx-sev}) ) that 
\begin{equation}
\exp[A_m(F_{m-1},F_m)]= 
-{1 + (e^{H_m} -1) F_m \over y_m \, F'(x_m)  } =  
-{1 + (e^{H_m} -1) F_{m-1} \over y_m \, F'(x_{m-1})  }  
\label{A-exp}
\end{equation}
one gets for ${\cal B}^*$
\begin{equation*}
\begin{split}
{\cal B}^* = &
 \left({\rho_b - \rho_a \over  y \,F'(0)}\right)^a  
 \left({\rho_b - \rho_a  \over y \,  F'(y)}\right)^b  \times \\
 & \left( 1  + \sum_m {x_m(y-x_m)  \over y}W_m
+ \sum_{m_1>m_2}{ (y-x_{m_1})  (x_{m_1}-x_{m_2}) x_{m_2}  \over y} W_{m_1} W_{m_2} + ... \right)^{-1/2 }
\end{split}
\end{equation*} 
This expression coincides with the expected expression (\ref{claim3}). Therefore  this subsection  has established the validity of (\ref{res-neq}) in the case of several boxes.
\subsection{ {\bf A large number of boxes} }
Let us now  try to take the large $n$ limit of the above result.  
We consider that we have $n$ boxes, of equal length $L/n=\lambda y / n$,  and that the  field  $H_m$
in the $m$-th box is  given by
 $$H_m= H\left({m  \, y  \over n} - {y \over 2 n} \right) $$
where $H(x)$ is a smoothly varying function.
For simplicity we choose the boxes of equal lengths. Therefore  one has
$$x_m ={m \, y \over n} \  . $$
One then need to solve the equations (\ref{Fm-eq}) satisfied by the $F_m$'s. 
For large $n$ one can show by a direct calculation that the solution of these equations is given by 

\begin{equation}
F_m = F\left({m \, y \over n} \right)  + O \left({1 \over n^2  } \right) 
\label{Fm-sol}
\end{equation}

 where $F(x)$ is the solution of  (\ref{Feq}). (Note that from now on, $F(x)$ is the solution of (\ref{Feq}) when $H(x)$ is a smoothly varying function. This solution $F(x)$ is not identical to (\ref{Fx-sev}) which was  obtained for a piecewise $H(x)$. The difference is  at the origin of the correction of order  $O(n^{-2})$  in (\ref{Fm-sol}). This difference will lead to negligible terms anyway.)
One has from (\ref{Am-def})
\begin{equation*}
\begin{split}
A_m(F_{m-1},F_m) = - \log{y \over n} + \log \left({1 + F(x_m) (e^{H(x_m) }-1)   \over -F'(x_m) } \right)   & \\ 
- {y \over  2 n}  \, {F(x_m) H'(x_m) e^{H(x_m)} \over  1 + (e^{H(x_m)} -1)F(x_m )} &+ O\left({1 \over n^2}\right)  
\end{split}
\end{equation*} 
which can be rewritten using the fact that $F(x)$ is solution of (\ref{Feq})
\begin{equation*}
\begin{split}
A_m(F_{m-1},F_m) = - \log{y \over n} &+ \log \left({1 + F(x_m) (e^{H(x_m) }-1)   \over -F'(x_m) } \right)  \\ 
& - {y \over 2 n}  \left. \left(  \log \left({1 + F(x) (e^{H(x) }-1)   \over -F'(x) } \right)\right)'\right|_{x=x_m}  + O\left({1 \over n^2}\right) 
\end{split}
\end{equation*} 
Using then the Euler McLaurin formula to perform the sum (\ref{calA}), one finds that
\begin{eqnarray}
{\cal A}(\rho_a,\rho_b)=   -n \log{y \over n} +   \int_0^y  \log \left({1 + F(x) (e^{H(x) }-1)   \over -F'(x) } \right)  dx + O\left({1 \over n} \right) 
\end{eqnarray}
and this leads (\ref{ZL-star},\ref{calA}) to the term proportionnal to $\lambda$ in (\ref{res-neq}).
\\ \ \\ \ \\
One can also obtain the large $n$ estimate of $W_m$
$$W_m = -{y \over n}   \left. \left({F''(x)  \over F'(x)} \right)' \right|_{x=x_m}  $$
Then   by defining $W(x)$ by
$$W(x)= -    \left({F''(x)  \over F'(x)} \right)' $$
one can see that 
\begin{equation*}
\begin{split}
{\cal B}&  =  {(\rho_a - \rho_b)^{a+b} \over (-F'(0))^a (-F'(1))^b } \ \ \times \\
 & \left( 
 1  + \int_0^y  {x(y-x)  \over y}W (x) dx   
+ \int_0^1     dx  \int_x^1 dz { (y-z) (z-x) x  \over y}   W(z) W(x) + \cdots 
\right)^{-1/2} 
\end{split}
\end{equation*} 

Then if one $\varphi(x)$  is solution of
$$\varphi''(x) =  W(x)  \varphi(x) $$
 with $\varphi(y)=0$ and $\varphi'(y)=-1/y$
one has
 
\begin{equation*}
\begin{split}
\varphi(x)=  1-  {x \over y}& + \int_x^y dz   { (y-z) (z-x) \over y}  W(z) \\
&+\int_x^y dz   \int_z^y dz' {  (y-z') (z'-z )(z-x) \over y}   W(z) W (z') + ...  
\end{split}
\end{equation*} 
and one finds
$$
{\cal B} 
= \left({\rho_b - \rho_a \over F'(0)} \right)^{a}  \left({\rho_b - \rho_a \over F'(y)} \right)^{b}  \left(  \varphi(0)   \right)^{-1/2} 
$$
as claimed in (\ref{res-neq}).

\section{Equilibrium case}
Let us consider a one dimensional 
lattice gas   on a lattice of $L$ sites,  where each site $i  $ is occupied by an integer $n_i \ge 0$ number of particles. We assume that the interactions are short range and that at equilibrium the system is  homogeneous   in the bulk with   correlation functions   decaying  exponentially fast with the distance.
We would like to obtain an expression of  $G_L(h_1, \cdots, h_L)$   defined in (\ref{Z-def},\ref{G-def})  for a slowly varying field of the form (\ref{slow})
when both  $\lambda$  and $ L$ are much larger than the range  $\xi$ of the correlations between the occupation numbers $n_i$. 
\subsection{ The constant field case}
Let us discuss first the case of a constant field ($h_i=h$).
 If  $g( h)$ is the extensive part  of the free energy $G_L$ 
$$ g(h) = \lim_{L \to \infty}{ G_L(h,\cdots h) \over L}$$ 
one  expects that in the large $L$ limit
\begin{equation}
G_L(h) =  L  \, g(h) +  A^{\rm left} (h) + A^{\rm right} (h) +  O \Big(\exp[-L / \xi(h)] \Big)
\label{G-asymp}
\end{equation}
 where $  A^{\rm left} (h)$ and  $ A^{\rm right} (h)  $ represent  the  contributions of the left and 
right boundaries respectively  and $\xi(h)$ is the correlation length (in presence of the constant field $h$).
The form (\ref{G-asymp}) can be easily understood by the transfer matrix method, in particular  $\exp(-1/ \xi(h))$ is the ratio of the two largest eigenvalues of the transfer matrix.
These two contributions $  A^{\rm left}$ and  $ A^{\rm right}$ are not necessarily equal  as they   may differ if one imposes different boundary conditions at the two ends.
 
In  a constant field, one can also define the  average density $\langle n_i \rangle$ at site $i$ by
$$\langle n_i \rangle = {\partial \log Z_L \over \partial h_i}(h, \cdots h)
  $$
and the pair correlation function 
$$\langle n_i  n_j  \rangle_c = {\partial^2 \log Z_L \over   \partial h_i \, \partial h_j}(h, \cdots h)
$$
In the large $L$ limit, far from the boundaries (i.e. for $i \gg \xi(h)$ and $L-i \gg \xi(h)$),   the average density  $\langle n_i \rangle$ has a limit independent of $i$ 
\begin{equation}
\langle n_i    \rangle \to  g'(h)  \  ,
\label{lim1}
\end{equation}
 the pair correlation function 
$\langle n_i  n_j  \rangle_c$ depends only on the distance $j-i$
\begin{equation}
 \langle n_i  n_j  \rangle_c \to c_{j-i}(h)
\label{lim2}
\end{equation}
and  one has
\begin{equation}
 g''(h)=  \sum_{k=-\infty}^\infty c_{k}(h) \ .
\label{lim3}
\end{equation}

On the other hand, close to the left or to the right boundary,  i.e. as long as $i \sim  \xi(h)$ or $L-i \sim \xi(h)$ these quantities keep in general a dependence on $i$ even in the large $L$ limit. For example
\begin{equation}
 \langle n_i      \rangle-  g'(h)  \to a_i^{\rm left }(h)
 \ \ \ \ ; \ \ \ \langle n_{L-i}\rangle  - g'(h)     \to a_i^{\rm  right}(h)
\label{lim4}
\end{equation}
\subsection{ The slowly varying  field case} 
Now  for a slowly varying field  of the form (\ref{slow}),  when $\lambda  \gg \xi(h) $ (more precisely $\lambda \gg \max_i \xi(h_i)$) one expects  
\cite{BDGJL6,derrida-2007}
that  to leading order
 $$G_L(h_1, \cdots h_L) \simeq  \lambda  \int_0^y  g( H(x)) \  dx $$
where $y$ is defined in (\ref{y-def}).
This can be easily understood by cutting the system of length $L$ into  many subsystems of size $\lambda dx$ much larger than $\xi$ but much smaller than $\lambda$. In each of these subsystems
the field $h_i$ is essentially constant, and the free energies of these subsystems can simply be added.

As explained in Appendix \ref{annexe2}, the leading corrections to this formula
  when   $L$ and $\lambda $ are much larger than the correlation length $\xi(h)$ is
\begin{eqnarray}
 G_L(h_1, \cdots h_L) & \simeq  & \lambda  \int_0^y  g( H(x))  \ dx
 \   +  \  A^{\rm left} (H(0)) \  + \  A^{\rm right} \left(H\left( y                 \right)\right)  
\nonumber \\ 
&&  +  \nonumber 
{1 \over \lambda}  \left[  H'(0)  \ B^{\rm left}   (H(0))
+ H'\left( y                 \right) \    B^{\rm right} \left(H\left( y                 \right)\right)    \right.
 \\ &&  \ \ \ \   \left. + \int_0^y                  C( H(x))\  H'(x)^2 \  dx 
\right]   + 0 \left( {1 \over \lambda^2 } \right) \label{equilibrium}
\end{eqnarray}
  which is the result announced in (\ref{res-eq}) where
\begin{equation}
\label{B-def}
 B^{\rm left} (h) =  {g'(h) \over  24} + \sum_{i=1}^\infty \left(i- {1 \over 2} \right)  \, a_i^{\rm left }(h) 
 \ \ \ ; \ \ \  B^{\rm right} (h) = - {g'(h) \over 24} - \sum_{i=0}^\infty \left(i+ {1 \over 2} \right)  \,   a_i^{\rm right }(h)
\end{equation}
and
\begin{equation}
\label{C-def}
C(h)= -{1 \over 2} \sum_{k \ge 1} \  k^2 \  c_k(h) 
\end{equation}

\ \\
{\bf Remark:} for the same  system on  ring with periodic boundary conditions,  implying in particular that $ H\left({y } +x \right) = H(x) $, the boundary terms disappear and one gets
\begin{eqnarray}
 G_L(h_1, \cdots h_L) \simeq  \lambda  \int_0^y                  g( H(x))  \ dx + 
{1 \over \lambda}   \int_0^y                  C( H(x))\  H'(x)^2 \  dx
 + \cdots 
\label{period}
\end{eqnarray}
\ \\
Note that at order ${1 \over \lambda}$ the integral of $H'(x)^2$  in (\ref{equilibrium}) and in (\ref{period}) is nothing but the square of the gradient of the Ginsburg Landau theory.
\ \\ \ \\

\section{Conclusion}

In this paper we have obtained the first correction (\ref{res-neq}) to the large deviation  functional of the density  for the non equilibrium steady state of the SSEP
and compared it  with the corresponding term for equilibrium systems (\ref{res-eq}). Like in the  equilibrium case, this first correction  does not depend on the system size.
On the other hand 
in the non-equilibrium case (\ref{res-neq})  the correction has a non-local character, 
very much like the leading term. Our derivation is based on the knowledge of the steady state  measure, as given by the matrix ansatz.

An interesting question would be to  try to recover our  result by  the macroscopic approach:
in  the macroscopic  fluctuation theory \cite{BDGJL1,BDGJL2,BDGJL5},   the  large deviation functional of the density  is given by the contribution of  the optimal trajectory in the space of all the time dependent density profiles which produces a given deviation starting from the steady state profile. A natural question would be to try to calculate the correction by integrating over all the profiles in the neighborhood of this optimal profile. Such an approach was successful in understanding the first corrections to the large deviation function of the current \cite{ring-fss,open-fss}, and it would be of course interesting to see whether it  works as well for the deviations of the density. If this is the case, 
one could try to determine similar corrections for other models such as generalizations of the SSEP \cite{enaud,BGL,CGGR}. 

Recently, it has been noticed that the large deviation functional could exhibit phase transitions
\cite{bertini-phase-transition,haifa3,haifa2}.
 Whether the corrections calculated here would become singular at such phase transitions is another  question one could try to investigate.

It has also been shown that the  SSEP, in a non equilibriuml steady state, could be mapped by a non local change of variables onto a system at equilibrium \cite{Kurchan}. It would be interesting to know whether this transformation could be sued to  confirm our prediction (\ref{res-neq}) and establish a connection with (\ref{res-eq}).

\begin{acknowledgements}
We would like to thank Vincent Hakim for  helpful discussions.
\end{acknowledgements}

\appendix
\section{ Additivity formulae }
\label{annexe1}

It is known \cite{DEHP,evans,derrida-2007} and has been used in  several previous works \cite{DLS2001,DLS2002a}  that
the steady state measure of the SSEP with injection and removal rates $\alpha, \beta, \gamma, \delta$, as defined in the introduction, can be calculated by the matrix ansatz.
\cite{DEHP,evans}. 
The probability  of any microscopic configuration $\{n_1, \cdots \ n_L\} $ (with $n_i=0$ or $1$)
is given by
\begin{equation}
P(\{n_1, \cdots \ n_L\} ) = 
  {\langle W | X_1 X_2 ... X_L | V
\rangle \over \langle W | (D+E)^L |V \rangle }
\label{matrix}
\end{equation}
where each  matrix $X_i$ depends on the occupation  $n_i$ of site $i$
\begin{equation}
X_i =  n_i  D + (1 - n_i) E
\end{equation}
 and the matrices $D,E$ and the vectors   $ | V \rangle , \langle W |$ satisfy the following algebraic rules
\begin{eqnarray}
&& DE-ED= D+E \nonumber \\
&& \langle W | ( \alpha E - \gamma D) = \langle W| \label{algebra} \\
&&  ( \beta D - \delta E)| V \rangle = | V \rangle  \; . \nonumber
\end{eqnarray}
Given these algebraic rules, one can  define a family of left and right eigenvectors
 $\langle \rho_a,a|$ and $|\rho_b,b \rangle$  by 
\begin{equation}
\label{left}
 \langle \rho_a,a| \Big(\rho_a E - (1-\rho_a) D \Big) = a \ \langle \rho_a,a| 
\end{equation}
\begin{equation}
\label{right}
  \Big( (1-\rho_b) D - \rho_b E  \Big) |\rho_b,b \rangle = b \  |\rho_b,b \rangle  \  .
\end{equation}
The vectors    $ \langle W |$ and $|V \rangle$
which appear in (\ref{matrix},\ref{algebra})  are   examples of such eigenvectors
\begin{equation}
  \langle W| =\langle \rho_a,a| \ \ \ \ , \ \ \ \
 | V \rangle =
|\rho_b,b \rangle  
\label{V-W}
\end{equation}
when 
$ \rho_a = \alpha /(\alpha + \gamma)$,
$ \rho_b = \delta /(\delta + \beta )$
and 
$a= 1/(\alpha + \gamma)$ , $ b= 1/(\delta + \beta ) $
as in  (\ref{ra-rb},\ref{a-b-def}). 
\ \\ \ \\
Then for $ 0 < b < 1$   and $\rho_a > \rho_b$, one can  prove the following {\it key  additivity formula} 
\begin{equation}
 {\langle \rho_a,a| X_1 X_2 |\rho_b,b \rangle \over \langle \rho_a,a|
\rho_b,b  \rangle}  =
\oint {d \rho  \over 2 \pi i}{ (\rho_a - \rho_b)^{a+b} \over (\rho_a-\rho)^{a+b}
(\rho-\rho_b) }{\langle \rho_a,a| X_1  |\rho,b \rangle 
 \over \langle \rho_a,a|\rho,b  \rangle}{ \langle \rho,1-b|  X_2 |\rho_b,b
\rangle \over \langle \rho,1-b|\rho_b,b  \rangle} 
\label{key}
\end{equation}
where $X_1$ and $X_2$   are arbitrary polynomials of $D$'s and $E$'s and  the contour is such that $\rho_b < |\rho | < \rho_a$.
\\ \\ \\
{\bf Proof of (\ref{key})}:
Let us first  derive of the following identity
\cite{DLS2002a}
\begin{equation}
{\langle \rho_a,a| (D+E)^L  |\rho_b,b \rangle  \over
\langle \rho_a,a |\rho_b,b  \rangle} = {\Gamma(a+b+ L) \over \Gamma(a+b)
\ (\rho_a - \rho_b)^L}
\label{normalization}
\end{equation}
To do so one can notice that in the steady state of the SSEP, as defined in the introduction,  the average occupations satisfy
$$ \alpha - (\alpha+ \gamma) \langle n_1 \rangle=
\langle n_1  -  n_2 \rangle  = \cdots \langle n_i - n_{i+1} \rangle  = \cdots = 
( \beta + \delta)  \langle n_L \rangle - \delta$$
These $L$ equations which  express simply that in the steady state the current is conserved,  can be solved.  
From the solution 
 (\ref{ni}) 
one can see that
$$  \langle n_i- n_{i+1} \rangle ={(\rho_a- \rho_b) \over L + a+b-1} \ . $$
On the other hand using the matrix representation (\ref{matrix},\ref{algebra},\ref{V-W}) one has
\begin{eqnarray*}
  \langle n_i  -  n_{i+1} \rangle  =
{\langle \rho_a,a|(D+E)^{i-1}(DE-ED) (D+E)^{L-i-1}   |\rho_b,b \rangle  \over
\langle \rho_a,a |(D+E)^L\rho_b,b  \rangle}  \\  = 
{\langle \rho_a,a| (D+E)^{L-1}   |\rho_b,b \rangle  \over
\langle \rho_a,a |(D+E)^L\rho_b,b  \rangle}  
\end{eqnarray*}
These two identities give the recursion 
$$
{\langle \rho_a,a| (D+E)^{L-1}   |\rho_b,b \rangle  \over
\langle \rho_a,a |(D+E)^L |\rho_b,b  \rangle}  
= {(\rho_a- \rho_b) \over L + a+b-1}$$
which   establishes the veracity of (\ref{normalization}).

Now 
to prove   (\ref{key})  (as in \cite{derrida-2007}) one can  first notice that  the discussion can be limited to  $X_1$ and $X_2$ of the form
$$X_1 = [\rho_a E - (1-\rho_a) D]^{p_1} \ [  D +  E]^{q_1} $$
$$X_2 = [ D + E]^{p_2} \ [ (1-\rho_b) D - \rho_b E]^{q_2} $$
as  any polynomial in $D$'s and $E$'s can be written as a sum of such terms
(this is because $D$ and $E$ are linear functions of the operators $A$ and $B$ defined by $A=D+E$
and $B= \rho_a E - (1-\rho_a) D$ and that $AB-BA=A$.  Thus   word made up of $A$'s and $B$'s can be ordered as a sum of terms of the form  $B^{p_1} A^{q_1}$ or $A^{p_2} B^{q_2}$) 
Then  the left hand side of  (\ref{key}) becomes
\begin{equation}
\label{u1}
 {\langle \rho_a,a| X_1 X_2 |\rho_b,b \rangle \over \langle \rho_a,a|
\rho_b,b  \rangle}  = a^{p_1} b^{q_2} 
{\langle \rho_a,a| (D+E)^{{q_1+p_2}} |\rho_b,b \rangle \over \langle \rho_a,a|
\rho_b,b  \rangle}
\end{equation}
while the right hand side  of  (\ref{key}) becomes
\begin{equation}
\label{u2}
a^{p_1} b^{q_2} \oint {d \rho  \over 2 \pi i}{ (\rho_a - \rho_b)^{a+b} \over (\rho_a-\rho)^{a+b}
(\rho-\rho_b) }{\langle \rho_a,a| (D+E)^{q_1}  |\rho,b \rangle 
 \over \langle \rho_a,a|\rho,b  \rangle}{ \langle \rho,1-b|  (D+E)^{p_2} |\rho_b,b
\rangle \over \langle \rho,1-b|\rho_b,b  \rangle} 
\end{equation}
and the equality of  (\ref{u1}) and (\ref{u2}) follows from the expression (\ref{normalization}) and the Cauchy theorem.
This completes the derivation of (\ref{key}).
\\ \\ \\
{\bf First consequence  of (\ref{key})}:
\\
It is possible  to show directly from the algebra (\ref{matrix},\ref{algebra})
that
$$ {\langle \rho_a,a|  D |\rho_b,b \rangle \over \langle \rho_a,a|
\rho_b,b  \rangle} =   { b \rho_a + a   \rho_b \over  \rho_a - \rho_b} \ \ \ ; \ \ \  
 {\langle \rho_a,a|  E |\rho_b,b \rangle \over \langle \rho_a,a|
\rho_b,b  \rangle} =   { b(1-\rho_a) + a (  1-\rho_b) \over  \rho_a - \rho_b} $$
which becomes by replacing $\rho_a$ by $\rho$ and $a$ by $1-b$
$$ {\langle \rho,1-b|  D |\rho_b,b \rangle \over \langle \rho,1-b|
\rho_b,b  \rangle} =  b + {   \rho_b \over  \rho - \rho_b} \ \ \ ; \ \ \  
 {\langle \rho,1-b|  E |\rho_b,b \rangle \over \langle \rho,1-b|
\rho_b,b  \rangle} =  - b + {   1-\rho_b \over  \rho - \rho_b} $$
Therefore (\ref{key}) becomes after integration
\begin{small}
\begin{equation}
 {\langle \rho_a,a| X_0 D |\rho_b,b \rangle \over \langle \rho_a,a|
\rho_b,b  \rangle}  =
b {\langle \rho_a,a| X_0  |\rho_b,b \rangle \over \langle \rho_a,a|\rho_b,b  \rangle}
+ \rho_b 
  \left. { d \over d \rho} \left[ 
\left({\rho_a - \rho_b \over \rho_a - \rho}\right)^{a+b}
{\langle \rho_a,a| X_0  |\rho,b \rangle \over \langle \rho_a,a|\rho,b
\rangle} \right] \right|_{\rho=\rho_b}
\label{additivity1}
\end{equation}
\begin{equation}
 {\langle \rho_a,a| X_0 E |\rho_b,b \rangle \over \langle \rho_a,a|
\rho_b,b  \rangle}  = -
b {\langle \rho_a,a| X_0  |\rho_b,b \rangle \over \langle \rho_a,a|\rho_b,b  \rangle}
+ (1- \rho_b) 
  \left. { d \over d \rho} \left[
\left({\rho_a - \rho_b \over \rho_a - \rho}\right)^{a+b}
{\langle \rho_a,a| X_0  |\rho,b \rangle \over \langle \rho_a,a|\rho,b
\rangle} \right] \right|_{\rho=\rho_b}
\label{additivity2}
\end{equation}
\end{small} 

These last two formulae are exact and valid for all values of
$\rho_a,\rho_b,a,b$. (They have been  derived from (\ref{key})
under the condition that $\rho_a > \rho_b$ and $0 < b < 1$,
but as all expressions are rational functions of all their arguments,
they remain valid everywhere.)

From  (\ref{additivity2}) 
it is possible to show 
that
$\Phi(\mu,h)  $ defined by
\begin{equation}
\Phi(\mu,h) = { \langle W |   \exp[ (e^h D + E) \mu ]   | V \rangle  \over \langle W |  V \rangle }
\end{equation}
satisfies the following equation
$$
{d \Phi \over d \mu}= { b(1+ \rho_a (e^h-1)) + a (1+ \rho_b (e^h-1)) \over \rho_a-\rho_b}  \Phi + (1+\rho_b (e^h-1)) { d \Phi \over d\rho_b}  $$
This equation can be solved by the method of characteristics, which tells us that the solution is of the form
$$\Phi(\mu,h)={ (\rho_a - \rho_b)^{a + b}  \over (1 + \rho_b( e^h-1) )^b}  \  {\cal F}  \left(  (1+ \rho_b(e^h -1) )\exp[\mu (e^h-1) ]   \right) $$
The fact that $\Phi(0,h)=1$  determines the unknown function ${\cal F}$ and one gets 
\begin{eqnarray}
\Phi(\mu,h)  = \left( {(\rho_a- \rho_b) ( e^h -1) \over 1+ \rho_a (e^h-1)    -
\exp[\mu(e^h-1)]
( 1 + \rho_b(e^h-1)) 
 } \right)^{a+b} \exp[b \mu (e^h-1) ] 
\nonumber \\
\label{generating-function} 
\end{eqnarray}
(see  eq (3.7-3.10) of \cite{DLS2002a}). 
\ \\ \ \\
 This  expression  becomes singular as \mbox{ $\mu \to \mu_0$ }with 
$$ \mu_0={1 \over e^h-1}\log\left({1+ \rho_a(e^h-1) \over   1+ \rho_b(e^h-1)}\right)$$ and  by analysing the power law singularity one can get the asymptotic expression valid for large $L$
\begin{equation}
\dfrac{\langle W   |    (e^h D + E)^L   | V \rangle  }{ \langle W |  V \rangle  } \simeq  \dfrac{\Gamma(a+b+ L) \ (\rho_a-\rho_b)^{a+b} \  \mu_0^{-L-a-b}  }{  \Gamma(a+b) \ (1+\rho_a(e^h-1))^a \ (1+\rho_b(e^h-1))^b  }
\label{1box-large-L}
\end{equation}
\\ \ \\ \ \\ 
{\bf Second  consequence  of (\ref{key})}: 
\\
Another important consequence  which can be obtained by dividing (\ref{key}) by (\ref{normalization})
is the following additivity formula
\begin{eqnarray}
\label{add}
  \frac{\langle \rho_a,a|X_1 X_2|\rho_b,b\rangle}
        {\langle \rho_a,a| (D+E)^{L+L'}    |  \rho_b,b\rangle}
   =   {\Gamma(L+a+b) \; \Gamma(L'+1) \over \Gamma(L+L'+a+b)}
\oint_{\rho_b<|\rho|<\rho_a}   \frac{d\rho}{2i\pi} \ \times
 \ \ \ \ \ \ \ \ \ \   \ \ \ \ \
 \\ 
    \ \ \ \ \ \   
   \frac{(\rho_a-\rho_b)^{a+b+L+L'}}{(\rho_a-\rho)^{a+b+L}(\rho-\rho_b)^{1+L'}}
      \frac{\langle \rho_a,a|X_1|\rho, b\rangle}
          {\langle \rho_a,a| (D+E)^L |  \rho, b\rangle} \:
     \frac{\langle \rho,1-b|X_2|\rho_b,b\rangle}
          {\langle \rho,1-b|  (D+E)^{L'} | \rho_b,b\rangle} \  .
\nonumber
 \end{eqnarray}
which is the same as eq. (65) of \cite{derrida-2007} up to the prefactor which was wrong in  \cite{derrida-2007}  and which is corrected here.
This formula allows one to compute the properties of a lattice of $L+L'$ sites if one knows those of  two  systems of size $L$ and $L'$.
\ \\ \ \\
{\bf Third   consequence   of (\ref{key})}:
\\
Using (\ref{additivity2}) and (\ref{normalization})  one can write an exact recursion for $ Z_L$  defined in (\ref{Z-def})
\begin{equation}
Z_{L+1}= \left[1+ \rho_b  \ e^{h_{L+1}} + b  \ {\rho_a-\rho_b \over L+a+b} \   e^{h_{L+1}}  \right] Z_L + { (\rho_a-\rho_b) (1+  \rho_b \  e^{h_{L+1}}   )
\over L+a+b}  \ {d  Z_L  \over d \rho_b}  
\label{additivity5}
\end{equation}
We won't use this recursion relation in this paper, but we believe that it could be  an alternative  starting point to  recover the result (\ref{res-neq}) and possibly further corrections.
\newpage

\section{  Derivation of (\ref{equilibrium},\ref{B-def},\ref{C-def}) in the equilibrium case }
\label{annexe2}
 Let us consider a  site dependent field  $h_i$ with small variations 
$$z_i= h_i - h         $$ 
around a certain value $h  $.
One can  then expand  $G_L$ defined in (\ref{Z-def},\ref{G-def}) in powers of the $z_i$'s 
$$G_L(h_1, \cdots h_L)    = 
G_L(h  , \cdots h  )    + 
\sum_i z_i \,  \langle n_i\rangle  + 
  {1 \over 2}  \sum_{i , j}   z_i z_j  \ \langle n_i n_j \rangle_c 
+ O\left(z^3 \right)  $$
where $\langle . \rangle$ denotes an  average in the constant field $h  $.
Far from the boundaries i.e. when $ i \gg 1$ and  $L-i \gg 1$, the correlations become translational invariant
(because the system is at equilibrium)
\begin{equation}
\langle n_i \rangle = g'(h  )   \ \ \ \ ;
 \ \ \ 
\langle n_i  n_j \rangle_c = c_{j-i}(h  )
\end{equation}
and
$$g''(h)= \sum_{k=-\infty}^\infty c_k(h) \ . $$
One can rewrite $G_L(h_1, \dots h_L)$ as 
\begin{eqnarray}
\nonumber
 G_L(h_1, \cdots h_L)    = 
G_L(h  , \cdots h  )    + g'(h  )  \sum_i z_i  + 
{c_0(h  )   \over 2 } \sum_i  z_i^2  +  \sum_{k \ge 1}   c_k(h  ) \,  \sum_{i=1 }^{L-k}   z_i \,  z_{i+k}  \\
+\sum_i z_i \Big(\langle n_i\rangle   -   g'(h  ) \Big)+
{1 \over 2} \sum_{i , j}   z_i z_j \Big(\langle n_i n_j \rangle_c -c_{j-i}(h  )\Big)
+ O\left(z^3 \right) 
\nonumber
\\ 
\label{exp2}
\end{eqnarray}
and using the fact  (wich follows from     (\ref{exp2}) by looking at the term proportional to $L$  when all the $h_i$'s are equal)  that
$$g(h_i)= g(h  ) + z_i\,  g'(h  )  + {z_i^2 \over 2}   \sum_{k =-\infty}  ^\infty  c_k(h  )+ O \left(z^3 \right)$$
one gets
\begin{eqnarray}
\label{GL5}
 G_L(h_1, \cdots h_L) &  - &  \sum_i g(h_i)  =     G_L(h, \cdots h)  - L g(h)
 \\ \nonumber  & & +  \sum_{k \ge 1}   c_k(h  )  \left[ \sum_{i=1 }^{L-k}  \left( z_i z_{i+k}-{z_i^2 + z_{i+k}^2 \over 2} \right)  -{1 \over 2}  \sum_{i= 1}^k z_i^2
- {1 \over 2}  \sum_{i= L-k+1 }^L z_i^2   \right]
\\
 & & +\sum_i z_i \Big(\langle n_i\rangle   -   g'(h  ) \Big)+
 {1 \over 2}  \sum_{i , j}   z_i z_j \Big(\langle n_i n_j \rangle_c -c_{j-i}(h  )\Big)
+ O\left(z^3 \right)  \ .\nonumber
\end{eqnarray}
\\ \ \\
In  the large $L$ limit, the correlation functions, near the boundaries,  have a limit which is not translational invariant
$$
 \langle n_i      \rangle-  g'(h  )  \to a_i^{\rm left }(h  )
 \ \ \ \ ; \ \ \ \langle n_{L-i} \rangle - g'(h  )      \to a_i^{\rm  right}(h  )
$$
 whereas
$$\langle n_i  \, n_j       \rangle_c-  c_{j-i}(h  )   \to b_{i,j}^{\rm left }(h  )
 \ \ \ \ ; \ \ \ 
\langle n_{L-i}  \, n_{L-j}       \rangle_c-   c_{j-i}(h  )     \to b_{i,j}^{\rm right }(h  )
$$
One then should have 
\begin{equation}
{d  A^{\rm left} (h) \over dh} = \sum_{i=1}^{\infty} a_i^{\rm left }(h) \ \ \ ;  \ \ \ 
{d  A^{\rm right} (h) \over dh} = \sum_{i=0}^{\infty} a_i^{\rm right }(h)
\label{a1}
\end{equation}
\begin{equation}
{d  a_i^ {\rm left } (h) \over dh} = \sum_{j} b_{i,j}^{\rm left }(h) - \sum_{k \ge i} c_{k}(h) \ \ \ ; \ \ \ 
{d  a_i^ {\rm right } (h) \over dh} = \sum_{j} b_{i,j}^{\rm right }(h) - \sum_{k \ge {i+1}} c_{k}(h)
\label{a2}
\end{equation}
 so that using  (\ref{G-asymp}) and the fact that $c_k(h)=c_{-k}(h)$ 
$${d^2  A^{\rm left} (h) \over dh^2} = \sum_{i\ge 1,j \ge 1}  b_{i,j}(h) - \sum_{k \ge 1}  k \, c_{k}(h)
 \ \ \  ; \ \ \  {d^2  A^{\rm right} (h) \over dh^2} = \sum_{i\ge 0,j \ge 0}  b_{i,j}(h) - \sum_{k \ge 1}  k\,  c_{k}(h)$$

For large $L$ this becomes
\begin{eqnarray*}
 G_L(h_1, \cdots h_L) &  - &  \sum_i g(h_i)  =   A^{\rm left}(h) + A^{\rm right}(h)
 \\  & & +  \sum_{k \ge 1}   c_k(h  )  \left[ \sum_{i=1 }^{L-k}  \left( z_i z_{i+k}-{z_i^2 + z_{i+k}^2 \over 2} \right)  -{1 \over 2}  \sum_{i= 1}^k z_i^2
- {1 \over 2}  \sum_{i= L-k+1 }^L z_i^2   \right]
\\
 & & +\sum_i z_i \Big(\langle n_i\rangle   -   g'(h  ) \Big)+
 {1 \over 2}  \sum_{i , j}   z_i z_j \Big(\langle n_i n_j \rangle_c -c_{j-i}(h  )\Big)
+ O\left(z^3 \right)
\end{eqnarray*}
which can be rewritten, up to terms of  third order  in the $z_i$'s 
\begin{eqnarray}
\label{G-eq}
 G_L(h_1, \cdots h_L) - \sum_i g(h_i) =  
 \   D^{\rm left} + D^{\rm right}-  \sum_{k \ge 1}   c_k(h_i)  \left[ \sum_{i=1 }^{L-k} { ( h_i- h_{i+k})^2 \over 2} \right]
\end{eqnarray}
 where
\begin{eqnarray*}
  D^{\rm left}  =  A^{\rm left} (h_1) - {1 \over 2} \sum_{k \ge 1}   c_k(h_1)   \left[ \sum_{i=1 }^k (h_i-h_1)^2   \right]
+\sum_i (h_i-h_1)  \, a_i^{\rm left }(h_1)
\\
 + {1 \over 2} \sum_{i , j}   (h_i-h_1) (h_j-h_1)\,   b_{i,j}^{\rm left }(h_1)
+ O\left(z^3 \right)
\end{eqnarray*}
and 
\begin{eqnarray*}
  D^{\rm right}  =  A^{\rm right} (h_L) - {1 \over 2} \sum_{k \ge 1}   c_k(h_L)   \left[\sum_{i=1 }^k (h_{L+1-i}-h_L)^2   \right]
+\sum_i (h_{L-i} -h_L)\,   a_i^{\rm right}(h_L)
\\
 + {1 \over 2} \sum_{i , j}   (h_{L-i}-h_L) (h_{L-j}-h_L) \,  b_{i,j}^{\rm right}(h_L)
+ O\left(z^3 \right)
\end{eqnarray*}
 All the differences $h_i-h_j$ which appear in (\ref{G-eq}) are between nearby sites $i,j$. 
Under this form,  the differences $h_i- h_j $ between remote sites do not need to be small. 
 In what follows we will assume that  (\ref{G-eq})  remains true   as long as these differences $h_i-h_j$  remain small  for nearby sites (i.e. for $|i-j| \ll \lambda)$  even if these differences could be large when $|i-j|  \sim \lambda$).

Now for a slowly varying field of the form (\ref{slow}), with $L= y \lambda$ as in  (\ref{y-def}) one can evaluate the different terms using the   Euler Mac Laurin formula 
$$\sum_{i=1}^L g(h_i) \simeq  \lambda  \int_0^y                 g(H(x)) dx   - { H'(y                )g'(H(y                )) - H'(0)  g'(H(0)) \over 24 \lambda } $$
\begin{eqnarray*}
  D^{\rm left}  \simeq   A^{\rm left} (H(0)) 
+ {H'(0)  \over \lambda} \sum_{i \ge 1}\left(  i -{1\over 2} \right)  \, a_i^{\rm left }(H(0))
\end{eqnarray*}
\begin{eqnarray*}
  D^{\rm right}  \simeq   A^{\rm right} \left(H\left(y                \right)\right) 
- {H'\left(y                \right)  \over \lambda} \sum_{i \ge 0} \left( i  +{1\over 2} \right) \, a_i^{\rm right }\left(H\left(y                \right)\right)
\end{eqnarray*}
$$
 \sum_{k \ge 1}   c_k(h_i)  \left[ \sum_{i=1 }^{L-k} { ( h_i- h_{i+k})^2 \over 2} \right] \simeq {1 \over  2\lambda} \int_0^{y               } \sum_{k \ge 1} k^2\,  c_k(H(x))\,  H'(x)^2 dx $$

\end{document}